\documentclass[a4paper,UKenglish,cleveref,pdftex]{lipics-v2019}

\newif\ifreport
\reporttrue

\usepackage{amssymb}
\usepackage{amstext}
\usepackage{amsmath}
\usepackage{amsthm}
\usepackage{multicol}
\usepackage{url}
\usepackage{graphicx}
\usepackage{hyperref}
\usepackage[capitalise]{cleveref}

\graphicspath{{./fig/}}

\newcommand\ADEC{\textsf{adec}}
\newcommand\AENC{\textsf{aenc}}
\newcommand\SDEC{\textsf{sdec}}
\newcommand\SENC{\textsf{senc}}
\newcommand\AESDEC{\textsf{sdec}}
\newcommand\AESENC{\textsf{senc}}
\newcommand\ATTACKER{\textsf{attacker}}

\newcommand\BITSTRING{\textsf{Bitstring}}
\newcommand\CLIENT{\textsf{Client}}

\newcommand\ELEM{\textsf{G}}

\newcommand\EVENT{\textsf{event}}

\newcommand\FINGERPRINT{\textsf{fingerprint}}
\newcommand\FORALL{\textsf{forall\ }}
\newcommand\FP{\textsf{fp}}
\newcommand\HASH{\textsf{hash}}
\newcommand\FUN{\textsf{fun\ }}

\newcommand\INJEVENT{\textsf{inj-event}}
\newcommand\IN{\textsf{in}}

\newcommand\IV{\textit{iv}}
\newcommand\KAES{\ensuremath{k_\textit{aes}}}
\newcommand\KDF{\textsf{kdf}}
\newcommand\KIV{\ensuremath{\textit{iv}_\textit{aes}}}
\newcommand\LETFUN{\textsf{letfun\ }}
\newcommand\LET{\textsf{let\ }}
\newcommand\MSG{\textsf{Msg}}
\newcommand\MSGKEY{\textsf{msgkey}}
\newcommand\MSGRECV{\textit{MsgRecv}}
\newcommand\MSGSENT{\textit{MsgSent}}
\newcommand\NEW{\textsf{new}}
\newcommand\NONCE{\textsf{Nonce}}

\newcommand\OUT{\textsf{out}}
\newcommand\PK{\textsf{pk}}
\newcommand\PRINCIPAL{\textsf{Principal}}
\newcommand\PRIVKEY{\textsf{PrivKey}}
\newcommand\QUERY{\textsf{query }}
\newcommand\REDUC{\textsf{reduc\ }}
\newcommand\SHAREDKEY{\textsf{SharedKey}}
\newcommand\SESSIONKEY{\textsf{SessionKey}}

\newcommand\SK{\textsf{sk}}

\ccsdesc[500]{Security and privacy~Security protocols}
\ccsdesc[500]{Security and privacy~Logic and verification}
\ccsdesc[500]{Security and privacy~Formal security models}

\nolinenumbers %
\hideLIPIcs  %

\title{Automated Symbolic Verification of Telegram's~MTProto~2.0}

\author{Marino Miculan}
{Department of Mathematics, Computer Science and Physics, University of Udine, Italy}
{marino.miculan@uniud.it}{0000-0003-0755-3444}{Supported by the Italian MIUR  project PRIN 2017FTXR7S \emph{IT MATTERS (Methods and Tools for Trustworthy Smart Systems)}.}

\author{Nicola Vitacolonna}
{Department of Mathematics, Computer Science and Physics, University of Udine, Italy}
{nicola.vitacolonna@uniud.it}{0000-0002-7505-5124}{}

\authorrunning{M.~Miculan, N.~Vitacolonna}
\Copyright{Marino Miculan, Nicola Vitacolonna}

\keywords{Specification, Verification and Synthesis; Security protocols; Practical verification; Privacy; Formal methods.}

\begin{document}

\maketitle

\begin{abstract}
 \ifreport
 \emph{MTProto~2.0} is a suite of cryptographic protocols for instant messaging at the core of the popular Telegram messenger application.
 \fi
  In this paper we analyse MTProto~2.0 %
  \ifreport
  \else, a suite of cryptographic protocols for instant messaging at the core of the popular Telegram messenger application,
  \fi
  using the symbolic verifier ProVerif.
  We provide fully automated proofs of the soundness of
  MTProto~2.0's authentication, normal chat, end-to-end encrypted chat,
  and rekeying mechanisms with respect to several security properties,
  including authentication, integrity, secrecy and perfect
  forward secrecy; at the same time, we discover that the rekeying protocol is vulnerable to an \emph{unknown key-share} (UKS) attack.
  We proceed in an incremental way: each protocol is examined in isolation, relying only on the guarantees provided by the previous ones and the robustness of the basic cryptographic primitives.
  Our research proves the formal correctness of MTProto~2.0 w.r.t.~most relevant security properties, and
  it can serve as a reference for implementation and analysis of clients and servers.
\end{abstract}

\section{Introduction}
Telegram is a very popular instant messaging application, with more than 500 million active users  (as of March 2021).
\ifreport
Besides user-to-user and group communication, \emph{channels} are widely adopted by newspapers, financial institutions, and even government agencies for broadcasting official news, in particular during emergency situations.
Telegram provides also an open API allowing third parties to offer (possibly commercial) services by means of \emph{bots}.

\fi
\ifreport
At the core of this ecosystem
\else
At its core
\fi
lies \emph{MTProto~2.0} \cite{Telegram-MTProto}, a suite of crypto\-graphic protocols designed for implementing fast, scalable, \emph{secure} message exchange
without relying on the security of the underlying transport protocol.
To this end, MTProto is composed by several (sub)protocols handling
the initial authentication of the clients with the creation of shared
keys between clients and server,
the creation of session keys between two clients for end-to-end encryption in secret chats, the rekeying of secret chats, and of course the encryption of all messages, before being transmitted over a (possibly insecure) network.

Although MTProto 2.0 is completely open and %
client's code is open-source,
Telegram's security model has received wide criticism.
First and foremost, the choice of the non-standard, \emph{ad hoc} protocol and encryption scheme has been objected, because the lack of scrutiny could expose the system to vulnerabilities, potentially undermining its security \cite{Kobeissi:2017}.
Moreover, all messages (even those of ``secret chats'') pass through
(a cloud of) proprietary, closed-source servers,
where they can be stored for any amount of time.
This architecture is convenient for users, who can access and synchronise their messages from several devices and send and receive messages also when the
peer is not present, but from a security perspective it means that the server must be considered as an \emph{untrusted} party.

This situation raises the need for a practical verification of the MTProto 2.0 protocol suite.
However, in spite of the criticisms above, most research has focused on the previous version MTProto~1.0, deprecated since December 2017.
To our knowledge, no formal, in-depth verification of MTProto 2.0 has been carried out so far.
This is the scope of this work.

In this paper we formalise and analyse MTProto~2.0 using ProVerif \cite{Blanchet:2016}, a state-of-the-art symbolic cryptographic protocol verifier based on the Dolev-Yao model.
We provide a fully automated proof of the soundness of
MTProto~2.0's protocols for authentication, normal chat, end-to-end encrypted chat,
and rekeying mechanisms with respect to several security properties,
including authentication, integrity, secrecy and perfect forward secrecy.
These properties are verified also in presence of malicious servers and clients.
On the other hand, we discover that, in principle, the rekeying protocol is vulnerable to an \emph{unknown key-share} (UKS) attack%
: a malicious client $E$ can induce two honest clients $A, B$ to believe they share two secret keys with $E$, and instead they share the same key between themselves only.

In order to prove these results we proceed in an incremental way.
Each protocol of the suite is examined in isolation, specifying which guarantees it requires from previous protocols and which ones it provides;
separation between protocols is guaranteed by the typing discipline enforced on messages%
\ifreport: ``out of sequence'' messages are  discarded\fi.
For each protocol we provide its formalisation in ProVerif%
\ifreport's specification language (the applied $\pi$-calculus)\fi%
, the formalisation of its security properties, and the results of the formal verification.

This approach allows us to cope with the complexity of the suite
and to isolate the security properties required on the underlying message encryption scheme.
Namely, the only assumption we make is that the latter is an \emph{authenticated encryption scheme}, guaranteeing both integrity of ciphertext (INT-CTXT) and indistinguishability of chosen plaintext (IND-CPA).
These properties are difficult to prove in a \emph{symbolic} model like ProVerif's, but can be proved in a \emph{computational} model, e.g.~using tools like CryptoVerif or EasyCrypt \cite{blanchet2007cryptoverif,barthe2013easycrypt}.
This assumption may appear strong, especially considering that Telegram has been widely criticized for its design choices%
, and vulnerabilities have been found in MTProto 1.0; however, none of these attacks have been replicated on MTProto 2.0.

In this paper we focus on the symbolic verification of MTProto~2.0 with respect to an unbounded number of sessions, leaving the analysis of the encryption scheme in the computational model to future work.
Proving the logical correctness of a protocol under a fairly general threat model is very important:
if a protocol is proved to be sound at a symbolic level then one can focus on searching vulnerabilities in the ``lower'' part of the protocol stack, among the chosen cryptographic functions and other implementation choices.
If a cryptographic primitive is found to be weak, it can be substituted with a stronger one, without invalidating the symbolic analysis.

Besides the relevance for Telegram's users, our formalisation can serve as a reference documentation for the implementation of clients and servers.
\vfill

\noindent
\emph{Synopsis.}
In \cref{sec:related-work} we recall previous related work.
The security model adopted is described in \cref{sec:security-model}.
In \cref{sec:mtproto} we recall the structure of MTProto 2.0, whose protocols are analyzed in the subsequent sections:
initial authorization key creation (\cref{sec:authorization}),
key exchange for secret chats (\cref{sec:secret-chat}),
rekeying in secret chats (\cref{sec:rekeying}).
Conclusions and directions for future work
are in \cref{sec:concl}.

We assume the reader confident with ProVerif; for an introduction we refer to \cite{Blanchet:2016} and several tutorials online. The code is available at
\url{https://github.com/miculan/telegram-mtproto2-verification}.

\section{Related Work}
\label{sec:related-work}
All the published research on MTProto that we are aware of, as well as most online articles, refer to the now deprecated MTProto~v1.0 and do not directly apply to the current MTProto~2.0, deployed in Telegram clients as of v4.6 (December 2017).

Arguably, the closest work to ours is~\cite{Kobeissi:2017}, where the Signal protocol is formalised in ProVerif and MTProto v1.0 is also briefly discussed, but not at the formal level.
The Signal protocol has been studied rigorously in \cite{frosch2016secure,cohn2017formal,rosler2018more}.

Several issues have been pointed out in MTProto v1.0.
Its encryption scheme added a random padding to the message prior to encryption but after the \emph{msg\_key} was computed, leading to a couple of theoretical~CCA attacks~\cite{Jakobsen:2016}.
Earlier versions of the protocol did not provide forward secrecy, and message sequence
numbers were managed by the server, so that a malicious server could
easily perform replay attacks~\cite{Jakobsen:2015}.
Another replay attack was discovered in Android's Telegram
client~v3.13.1~\cite{Susanka:2017}, where the same message could be
accepted twice by a client after 300 more messages had been sent. This
was due to a flaw in the implementation, which did not abide by
Telegram's Security Guidelines for Client Developers;
it was fixed in Telegram~v3.16.

The above mentioned issues were addressed in MTProto~2.0, whose developers claim to be IND-CCA and INT-CTXT secure and to provide~perfect forward secrecy for secret chats. There is currently no known proof of such claims, but there are no known attacks on MTProto's encryption, either.

A theoretical MitM attack to MTProto~1.0 has been described in \cite{Rad:2015}.
As we will see in \cref{sec:secret-chat}, the DH exchange used to establish a shared key before initiating a secret chat is not authenticated by the two ends.
Clients are supposed to verify
a hash of the shared secret through an external secure channel. In
MTProto~v1.0, the first 128 bits of the SHA1 of the key are used as
the fingerprint. A malicious server might social engineer two
clients to both initiate a conversation with each other; since the
server forwards all the messages, it might act as a MitM and try to
find two keys whose fingerprints coincide, using a birthday attack
and approximately $2^{65}$ computations~\cite{Rad:2015}.
In MTProto~2.0 (starting with the
``layer 46'' of secret chat protocol), the fingerprint is
288 bits long (additional 160 bits are extracted from the prefix of
the SHA256 of the key), thus making this MitM attack
likely infeasible.
\vspace{-1ex}

\section{Security Model}
\label{sec:security-model}
We model Telegram protocols in ProVerif~\cite{Blanchet:2016}, which is a \emph{symbolic} cryptographic verifier. Protocols and security properties are specified in a variant of the \emph{applied $\pi$-calculus}, a formalism designed for representing cryptographic processes, and translated into a Horn theory. Cryptographic primitives are represented by means of a suitable term theory, by means of constructors and reduction rules or equations; thus, cryptographic primitives are modeled as ``perfect''; in particular, there is no way to recover a plaintext or a key from the ciphertexts, and the ciphertext is not forgeable.

Following this approach, in our model we consider the message encryption scheme used in MTProto 2.0 essentially as an authenticated encryption scheme, abstracting from its actual implementation.
Formally, MTProto's symmetric encryption primitive is governed by the following reduction rule:
\begin{align*}
\FUN& \SENC(\BITSTRING, \SHAREDKEY, \NONCE): \BITSTRING \\
    & \REDUC \FORALL m\colon \BITSTRING, k\colon \SHAREDKEY, n \colon \NONCE; \\
    & \quad\SDEC(\SENC(m,k,n),k) = m.
\end{align*}

\ifreport
\noindent A
\else
Due to space limitations, a
\fi
detailed verification of these cryptographic functions in the computational model is left to future work.
\medskip

\noindent\textbf{\textsf {Threat Model.}}
We adopt the classical symbolic Dolev-Yao model \cite{Dolev:1983}, which is the one used by ProVerif. More specifically, we assume that all messages are transmitted over a public network, and that an active intruder can intercept, modify, forward, drop, replay or reflect any message. Besides, we assume that
an attacker may also exfiltrate secret data, such as pre-shared keys, during or after the execution of a protocol. As mentioned above, we assume that encrypted messages are unbreakable unless the key becomes available to the attacker. The model for hash functions is also quite strong, being close to the random oracle model. Other threats, such as timing and guessing attacks, are beyond the scope of our model.

All messages exchanged by clients pass through Telegram's servers.
Hence, such servers can access the plaintext of cloud-based chats and the ciphertext of secret chats.
Servers are also responsible for choosing the Diffie-Hellman parameters used to derive the long-term authorization keys for the clients.
Therefore, a server should not be considered as trusted.
\medskip

\noindent\textbf{\textsf {Security Goals.}}
Each part of MTProto has different security goals, as we will see later on.
In general, we will consider the following, informally stated, goals:
\begin{description}
  \item[Secrecy:]
  if a message $m$ is exchanged in a session~$S$ between two honest principals $A$ and~$B$ then $m$ is secret \ifreport (i.e., known only to $A$ and~$B$) \fi unless an attacker can break some cryptographic construction or recover the encryption keys before or during~$S$.
  \item[Forward secrecy:]
  secrecy of message $m$ is preserved even if the attacker recovers the encryption keys after $S$ is completed.
  \item[Authentication:]
  if $B$ receives $m$ which is supposed to come from~$A$, then it was really sent by $A$.
  \item[Integrity:]
  if~$m$ is sent from~$A$ to~$B$ then $B$ receives~$m$ and not some forged $m'\neq m$ instead.
\end{description}

\section{The MTProto 2.0 Protocol}
\label{sec:mtproto}
\ifreport
In this section we give a high-level overview of MTProto~2.0. For a deeper (albeit informal) description, we refer to the official web page \cite{Telegram-MTProto}.
\fi

MTProto 2.0 \ifreport\else \cite{Telegram-MTProto} \fi is a client/server protocol suite designed for accessing a (MTProto) server\footnote{Actually, Telegram employs a network (a ``cloud'') of servers in multiple data centers, spread worldwide for scalability and availability. However, for our aims, we can consider this network as a single server.}
from applications running on desktop computers or mobile devices, through an insecure network.
\begin{figure}[t]
  \ifreport
  \centering
  \includegraphics[width=0.75\columnwidth]{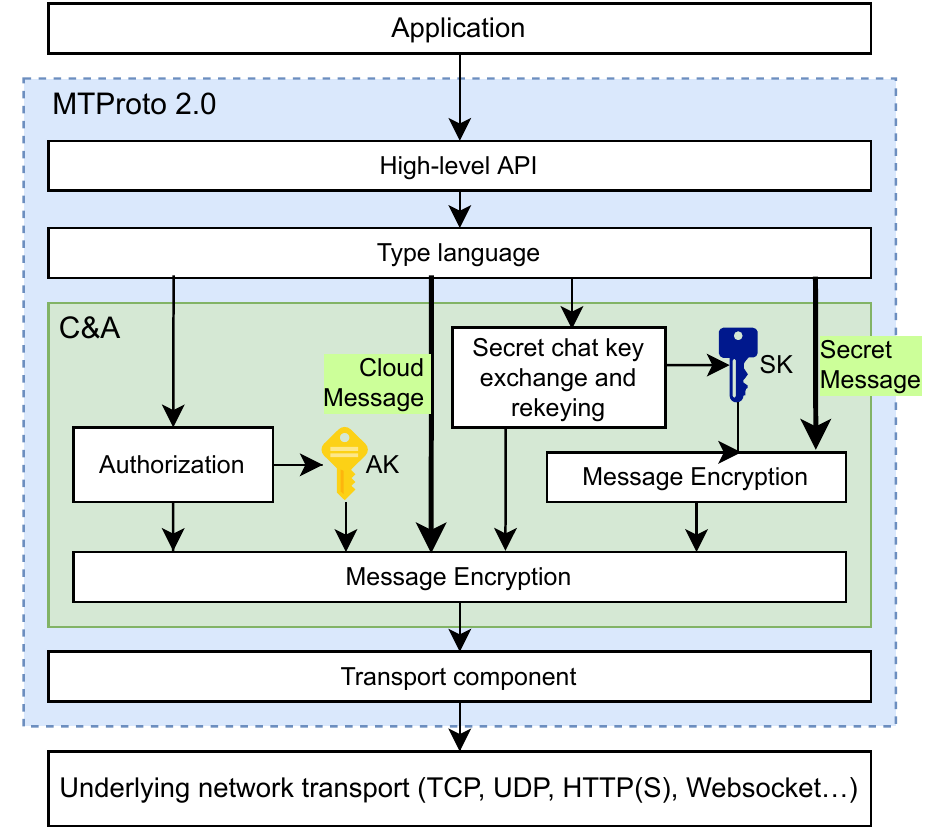}
  \else
  \begin{minipage}[b]{\textwidth}
  \fi
    \caption{\looseness=-1 
      The MTProto 2.0 suite.
      The subject of the present work is the ``Cryptography and Authorization'' (C\&A) component.
      AK (yellow key) is the Authorization Key, established once at the first run.
      SK (blue key) is the  Secret Chat Session Key, established at the beginning of each secret chat (and changed often).
      Cloud messages are encrypted only from client to server (and vice versa), with the AK.
      Secret messages are encrypted twice: with the SK, and then with the AK.
      In this picture the Message Encryption module is duplicated, but actually it is the same, with different keys.
    }
  \label{fig:stack}
  \ifreport
  \else
  \end{minipage}\hfill
  \includegraphics[width=0.5\columnwidth]{MTProto2-stack}
  \fi
\end{figure}
This suite can be divided into three main components (see \cref{fig:stack}):
\begin{description}
  \item[High-level API and type language:] defines how API queries and responses are converted to binary messages. This component fits OSI layers 7 (application) and 6 (presentation).
  \item[Cryptographic and authorization components:] defines how applications are authenticated with the server, and messages are encrypted before being transmitted through the transport protocol. These components fit OSI layers 5 (session) and 4 (transport).
  \item[Transport component:] defines how the client and the server actually exchange messages, via existing transport protocol such as
  UDP, TCP, HTTP, HTTPS, Websocket, etc.. Note that also insecure, connectionless protocols are supported.
\end{description}
We focus on the component handling cryptographic transformations and authorization.
This component can be divided further into the following modules:
\begin{description}
  \item[Authorization:] this module provides the functionalities for the initial client authorization and server authentication. It is called on the first run of the application, for deriving the \emph{authorization key} (AK), a long-term ``master'' secret shared with the server only.
    In order to establish the authorization key, this module executes a cryptographic protocol (basically a DH exchange) with the server.
    We will analyse this protocol in \cref{sec:authorization}.

  \item[Secret chat key exchange and rekeying:] this mod\-ule provides the functionalities for establishing a session shared secret key (SK) between two clients.
  It is executed once at the beginning of a secret chat and after 100 exchanged messages between the two parties (or over a week) for installing a new key.
  In both cases, this module executes a a Diffie-Hellman exchange with the peer client (through the server).
  We will analyse this protocol in \cref{sec:secret-chat,sec:rekeying}.

  \item[Message encryption:] All messages between client and server are encrypted with a symmetric cipher, using an ephemeral key derived from the AK.
  Messages in secret chats are encrypted also with an ephemeral key derived from the SK.
  The encryption scheme is the same, but it is use twice (with different keys) for messages in secret chats.
\end{description}
\smallskip
We stress that peer clients never communicate directly: messages always go through a server, where they are stored to permit later retrieval by the recipient. Cloud chat messages are kept in clear text, while secret chat messages are encrypted with the peers' session key, which should be unknown to the server.

\section{Authorization Keys}
\label{sec:authorization}%
On its first run, each Telegram client~$A$ must negotiate a long-term secret with a Telegram server~$S$.
Such an \emph{authorization key} is created through a Diffie-Hellman~(DH) exchange, it is never transmitted over the network, it is used for all subsequent communication between $A$ and~$S$, and it is almost never changed (basically, only when the client application is uninstalled and installed again).

\subsection{Informal description}
\label{sec:auth-informal}%
\ifreport
\begin{figure}[t]
  \centering
  \includegraphics[width=0.6\columnwidth]{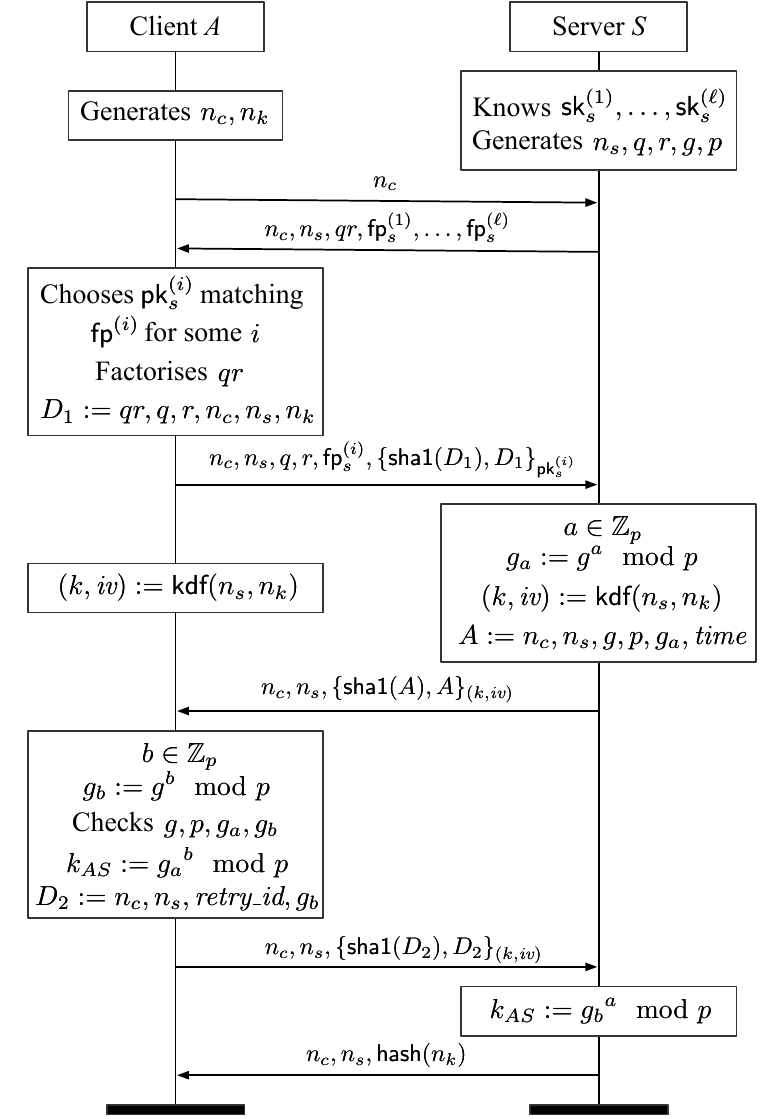}
  \caption{MTProto 2.0's Authorization Protocol.}
  \label{fig:auth-key-protocol}
\end{figure}
\else
\noindent
\begin{minipage}[b]{0.43\columnwidth}
\fi
The protocol consists of three rounds (\cref{fig:auth-key-protocol}):

\noindent\textbf{Round 1:} $A$ and~$S$ exchange a pair of randomly generated nonces $n_c$ and~$n_s$, which are sent along with all subsequent messages, both in plaintext and in the encrypted part of each message.
Such pair is used to identify a run of the protocol.
  \begin{enumerate}
    \item
    $A$ sends a nonce~$n_c$ to~$S$.
    \item
    $S$ replies with a message that contains $n_c$, a fresh nonce~$n_s$ generated by~$S$, a challenge $qr<2^{63}$ (to prevent DoS attacks against Telegram servers), which is the product of two primes $q$ and~$r$, and a list~$\FP_s^{(1)},\ldots,\FP_s^{(\ell)}$, with $\ell>0$, of the fingerprints
    of the public keys accepted by~$S$
    (computed as the 64~lower bits of the SHA-1 of the keys).
  \end{enumerate}

\noindent\textbf{Round 2:} $A$ decomposes $qr$ into its factors $q$ and~$r$,
  and retrieves\footnote{It is assumed that the server's public keys are known to the clients. In practice, they are usually embedded in Telegram's client apps. Of course, it is possible that a malicious client embeds a different key without the user to notice.} the public key~$\PK_s^{(i)}$ corresponding to some fingerprint~$\FP_s^{(i)}$.
  It also generates a new
\ifreport
\else
\end{minipage}\hfill
\includegraphics[width=0.55\textwidth]{MTProto2-auth}
\fi
  256-bit random secret nonce~$n_k$, which, together with the public~$n_s$, is used by both parties to derive an ephemeral symmetric key~$k$ and initialization vector~$\textit{iv}$ by hashing some substrings of~$n_s$ and~$n_k$.
  These data are used for encrypting the subsequent messages.
  \begin{enumerate}\setcounter{enumi}{2}
    \item
    $A$ serializes $(qr,q,r,n_c,n_s,n_k)$ and encrypts with~$\PK_s^{(i)}$ both this serialized data and its hash; then, it sends $n_c,n_s,q,r,\FP_s^{(i)}$ along with the encrypted payload to~$S$.
    \item
    $S$ chooses DH parameters $g$ and~$p$ and computes $g_a=g^a \mod p$ for some random 2048-bit number~$a$.
    Then, $S$ serializes $(n_c,n_s,g,p,g_a,t)$, where $t$ is the server's current time, and encrypts both the serialized data and its hash using a temporary symmetric key~$k$ derived from $n_s$ and~$n_k$.
    It then sends $n_c,n_s$ along with the encrypted payload to~$A$.
  \end{enumerate}

\noindent\textbf{Round 3:}
After deriving~$(k,\textit{iv})$ using the same algorithm as the server, $A$ decrypts the received message and checks that the received DH parameters are safe (see below).
  Then $A$ computes $g_b=g^b \mod p$ for some random 2048-bit number~$b$ and derives the authorization key $k_{AS}={g_a}^b\mod p$.
  \begin{enumerate}\setcounter{enumi}{4}
    \item \looseness=-1
     $A$ serializes $(n_c,n_s,\textit{retry\_id},g_b)$, where $\textit{retry\_id}$ is 0 at the first attempt to send this message, and it is equal to a hash of the previous authorization key if the server later asks to renegotiate the key within the same session by generating a new~$b$ (the server might ask the client to do this for the failure of the uniqueness check performed at the end of the protocol---see next paragraph).
    $A$ hashes the serialized data and encrypts both the hash and the data with~$(k,\textit{iv})$.
    It then sends $n_c,n_s$ along with the encrypted payload to~$S$.
    \item
    $S$ derives the shared key $k_{AS}={g_b}^a\mod p$, then verifies that $k_{AS}$ is unique by comparing a hash of $k_{AS}$ to the hashes of all authorization keys known to the server.
    If the hash is unique, $S$ sends an acknowledgment $(n_c,n_s,\textsf{hash}(n_k))$ to~$A$, otherwise $S$ sends an error message.
  \end{enumerate}
All the encrypted messages include a SHA1 hash of the content.
According to Telegram's ``Advanced FAQ''~\cite{Telegram-Tech-FAQ}, such hashes are ``irrelevant for security'', so they are presumably used only an additional sanity check for the implementation, and not as a means of implementing an AEAD (authenticated encryption with additional data) scheme.

The client is required to check that both $p$ and~$(p-1)/2$ are prime, that $2^{2047}<p<2^{2048}$ and that $g$ generates a cyclic subgroup of prime order $(p-1)/2$.
Both parties must also verify that $1<g,g_a,g_b<p-1$.
Telegram also recommends that both the client and server check that $2^{2048-64}\le g_a,g_b\le p-2^{2048-64}$.
Such checks should prevent the use of small subgroups and malicious primes, but it has already been noted that they could be made optional if Telegram used standardized values~\cite{Kobeissi:2017}.

\subsection{Formalisation in ProVerif}
\label{sec:auth-formal}%
In the formalization of the protocol we made a few simplifications: we ignored \textit{time} (relevant only to correctly generate message identifiers in later communication, which we do not model because we consider messages abstractly), \textit{retry\_id} (thus modeling explicitly only a successful path of execution---failures are implicitly taken account of by running an unbounded number of sessions, some of which may not complete), the proof-of-work decomposition of~$qr$ (because DoS attacks are not in our threat model) and, more importantly, the SHA1 hashes inside the encrypted messages.
Including such aspects into a symbolic model would significantly increase the computation times required for the automated verification, but it would most likely not allow us to discover more vulnerabilities.
The latter simplification, in particular, does not affect ProVerif's results (but it improves verification times), because in the symbolic model encryption is already authenticated, so adding a MAC is not really needed; besides, that aligns with Telegram's statement of ``irrelevance'' of such tags for security.

For this protocol, public-key encryption is modelled in the standard way using a reduction of the form $\ADEC(\AENC(x,\PK(k)),k)=x$, where $\AENC()$ is the encryption function, $\ADEC()$ is the decryption function and $\PK(k)$ is the public key corresponding to private key~$k$.
\ifreport
Thus, the message encryption scheme is assumed to be a secure authentication encryption scheme.
\fi
We assume that honest parties behave as mandated by the protocol, except for the following misbehaviour: a client may fail to verify that the received DH parameters are good, as explained in \cref{sec:auth-informal}; and the server is allowed to reuse the same nonce~$n_s$ in different sessions. Misbehaving processes are executed in parallel with correct processes.

Weak DH parameters are modelled as in~\cite{Bhargavan:2017}.
The generation of (possibly weak) DH parameters is delegated to two parallel processes as explained below.
A weak calculation always returns the same element, thus conservatively modeling subgroups of size~1.
Each computation involving a weak group or a bad element is reduced to the same bad value.
It is worth stressing that other equalities that hold in groups are not modelled,
although this model is stronger than the Diffie-Hellman assumption, in the sense that is impossible to compute $g^{xy}$ given only generator $g$, $g^x$ and~$g^y$.

Process macros are interleaved with event markers (``$\EVENT\ P(\vec x)$'' for some predicate $P(\vec x)$), which can be used to check whether a certain point in a process is reachable---hence, whether a certain event has happened. That allows us to specify some stringent correspondences (see \cref{sec:auth-security}). Events are also used to signal when a secret is compromised. Leaking information is modelled via several short processes that run in parallel with the clients and the server. For instance, compromising the server's private key is achieved by running a process that receives the server's private key from the server through a private channel~$c_{\textit{priv}}$ and makes it accessible to the attacker through the public channel~$c$, recording the event:
\begin{align*}
\LET\ &\textsf{LeakRSAKey}() =\\
      & \IN(c_{\textit{priv}}, k\colon\PRIVKEY);\\
      & \EVENT\ \textsf{CompromisedRSAKey}(k);\\
      & \OUT(c,k).
\end{align*}
Similar parallel processes are executed to perform the following functions and emit the corresponding event:
\begin{itemize}
  \item
  $\textsf{CompromisedNonce}(n_k)$: leaking the secret nonce~$n_k$ generated by the client during a session;
  \item $\textsf{PostCompromisedNonce}(n_k)$:
  revealing $n_k$ after the protocol is completed;
  \item $\textsf{PostCompromisedRSAKey}(k)$:
  revealing the ser\-ver's private key~$k$ after the protocol is completed;
  \item $\textsf{PostCompromisedAuthKey}(k)$:
  revealing the authorization key~$k$ after the protocol is completed (to test forward secrecy);
  \item $\textsf{ForgedServerIdentity}(S)$:
  associating the server's identity~$S$ to a private key chosen by the attacker.
\end{itemize}
Other four parallel processes are responsible for generating a random or fixed server nonce $n_s$, respectively, and for generating good or bad DH parameters, respectively, on behalf of the server (by communicating with the server through private channels). In general, we found that using a higher number of relatively simple processes rather than a smaller number of parametrized processes with a more complex logic allows ProVerif to perform much better, and gives the user more flexibility in expressing queries.

\subsection{Security properties verification}
\label{sec:auth-security}
\subparagraph{Authentication.}
The protocol for generating an authorization key does not prevent an intruder to act as a MitM during a registration session between a client~$A$ and a server~$S$ and impersonate~$A$ in subsequent exchanges with~$S$.
In other words, the protocol does not guarantee the authentication of the client to the server---which is expected, as the server is willing to run the protocol with any entity.
This is formalized with the following query:
\begin{align*}
  &\QUERY n_c\colon\NONCE,n_s\colon\NONCE;\\
  &\EVENT(\textsf{ServerAcceptsClient}(n_c,n_s))
  \Rightarrow\EVENT(\textsf{ClientRequestsDHParameters}(n_c,n_s)),
\end{align*}
for which ProVerif can find a counterexample even when the protocol is run correctly and without any information leak.
The query asserts that, if the server accepts a client in a session identified by~$(n_c,n_s)$, then it was that client who started session~$(n_c,n_s)$~---which may not be the case, because an attacker can take over after the first round and replace the legitimate client.

Failing authentication should not adversely affect the outcome of a session (except that $A$ must possibly restart the protocol in a new session).
The only result the intruder could achieve is a negotiation of an authorization key with~$S$, unrelated to~$A$.
Vice versa, it is important that $A$ knows with certainty that she has engaged with~$S$ and not with an attacker.
Authentication of the server to the client is proved by ProVerif with following query:
\begin{align*}
&\QUERY\SK_s\colon\PRIVKEY,\;n_c,n_s,n_k\colon\NONCE,\;g,g_a\colon\ELEM\;\\
&\quad\INJEVENT(\textsf{ClientReceivesDHParameters}(n_c,n_s,n_k,g,g_a))\\
&\quad\Rightarrow\INJEVENT(\textsf{ServerSendsDHParameters}(n_c,n_s,n_k,g,g_a))\\
&\quad\quad\lor\EVENT(\textsf{CompromisedRSAKey}(\SK_s))\\
&\quad\quad\lor\EVENT(\textsf{CompromisedNonce}(n_k))\\
&\quad\quad\lor\EVENT(\textsf{ForgedServerIdentity}()).
\end{align*}
The query is interpreted as follows: unless $S$'s private key~$\SK_s$ is compromised before or during the session identified by~$(n_c,n_s)$, or the secret nonce~$n_k$ is leaked during the session, or the server's identity is forged to begin with (e.g., the client embeds a spoofed public key), if $A$ accepts Diffie-Hellman parameters $(g,g_a)$ in session~$(n_c,n_s)$ after sending $n_k$ then $A$ is sure that it was the server who accepted $n_k$ and sent~$(g,g_a)$ in session~$(n_c,n_s)$.
This holds even if the server reuses $n_s$ in different sessions, because only $S$ can decrypt~$n_k$ and derive the ephemeral key with which DH parameters are transmitted.
Note that, since authentication events are registered before $A$ verifies the values received from~$S$, for authentication it does not matter whether the client checks that DH parameters are strong or weak.
The last disjunct reflects the fact that Telegram, as is common for current messenger apps, is based on a Trust-On-First-Use (TOFU) authentication infrastructure.

\subparagraph{Secrecy and forward secrecy.}
The authorization protocol provides secrecy for so called ``cloud messages'', i.e, the messages subsequently exchanged between the client and the server, which are encrypted using the shared authorization key. Namely, ProVerif proves the following query:
\begin{align*}
\QUERY&\textsf{sk}_s\colon \PRIVKEY, k\colon \SHAREDKEY, n_k\colon \NONCE, m\colon\MSG;\\
      &\ATTACKER(m)\\
      &\Rightarrow \EVENT(\textsf{CompromisedRSAKey}(\textsf{sk}_s))\\
      &\quad\lor\EVENT(\textsf{ForgedServerIdentity()})\\
      &\quad\lor\EVENT(\textsf{CompromisedNonce}(n_k))\\
      &\quad\lor\EVENT(\textsf{ChecksDHParameters}(\bot))\\
      &\quad\lor\EVENT(\textsf{PostCompromisedAuthKey}(k)).
\end{align*}
In words, under the assumptions of our model, the secrecy of message~$m$ is guaranteed unless:
\begin{enumerate}
  \item the server's private key is compromised or its identity is forged before or during the session that establishes the shared authorization key, or
  \item the secret nonce~$n_k$ generated by the client is leaked during such a session (in particular, if the attacker learns~$n_k$ before the fourth message then it can compute $(k,iv)$ and act as a MitM to establish two distinct authorization keys with the client and the server, respectively), or
  \item the client fails to validate the DH parameters received from the server, or
  \item \label{it:authcompr} the authorization key is compromised at any later time.
\end{enumerate}
This result holds even if the server reuses the same nonce~$n_s$ in multiple sessions with different clients. Besides, the result is strict, in the sense that removing any event from the query above leads to a counterexample. Since in our formalization the private key of the server and the secret nonce $n_k$ are always leaked in a separate phase following the completion of the authorization protocol (so that the impact of such leakage of information can be formally assessed), we may also conclude that leaking the server's key or the secret nonce \textit{after} a session has been completed does not violate the secrecy of subsequent client-server communication encrypted with the authorization key.
Intuitively, this holds because, even if such post-compromise allows an attacker to decrypt all the messages exchanged during the definition of the authorization key, she cannot compute the authorization key knowing only $g$, $m$, $g_a$, $g_b$ (assuming that the DH problem is hard).
Note also that case \ref{it:authcompr} above means that there is no guarantee of forward secrecy for messages encrypted with an authorization key.

After two clients have negotiated their authorization keys with
a server, they may start to exchange messages
within so-called \emph{cloud-based chats}. Every such message is
encrypted by the sender using the sender's authorization key and
forwarded to the server, who deciphers it and re-encrypts it with
the recipient's authorization key. In this context, the server can
trivially read (and even modify) every message. The previous result shows that, under
the hypothesis that the server is trusted, communication can at
least be kept secret against an external attacker. Cloud-based
chats do not provide forward secrecy, though: if, at any time, the
authorization key of one of the clients is leaked then all the
messages exchanged by that client can be deciphered.

\subparagraph{Integrity.}
Key agreement, a basic property of Diffie-Hellman, can be proved by ProVerif, using the following query:
\begin{align*}
&\QUERY n_c,n_s,n_k\colon\NONCE, k,k'\colon\SHAREDKEY, \textsf{sk}_s\colon\PRIVKEY;\\
&\quad\EVENT(\textsf{ServerAcceptsAuthKey}(n_c, n_s, k))\\
&\quad\land\EVENT(\textsf{ClientAcceptsAuthKey}(n_c, n_s, k'))
\end{align*}
\begin{align*}
&\quad\Rightarrow k = k'\\
&\quad\quad\lor\EVENT(\textsf{ClientChecksDHParameters}(\bot))\\
&\quad\quad\lor\EVENT(\textsf{ForgedServerIdentity}())\\
&\quad\quad\lor\EVENT(\textsf{CompromisedNonce}(n_k))\\
&\quad\quad\lor\EVENT(\textsf{CompromisedRSAKey}(\textsf{sk}_s)).
\end{align*}
If client and server generate authorization keys $k$ and~$k'$ during the same session identified by~$(n_c,n_s)$, then they compute the same key, unless the client fails to validate the DH parameters, the server's identity is forged, or some secret is leaked during the session.

Conversely, if client and server end a run of the protocol agreeing on the same key in their respective sessions, then such sessions coincide, unless the client skips its mandatory checks. That is, the following query is proved by ProVerif:
\begin{align*}
&\QUERY n_c,n'_c,n_s,n'_s\colon\NONCE, k\colon\SHAREDKEY;\\
&\quad\EVENT(\textsf{ServerAcceptsAuthKey}(n_c, n_s, k))\\
&\quad \land \EVENT(\textsf{ClientAcceptsAuthKey}(n'_c, n'_s, k))\\
&\quad\Rightarrow (n_c=n'_c\land n_s=n'_s) \\
&\quad\quad\lor\EVENT(\textsf{ClientChecksDHParameters}(\bot)).
\end{align*}
The last disjunct accounts for the fact that, if the server sends the same bad DH parameters to a client and to the attacker (acting as a malicious client) in two separate sessions and the honest client does not reject them, then the client, the attacker and the server may end up with the same key.

\section{Secret Chats}
\label{sec:secret-chat}
An end-to-end encrypted chat between two clients $A$ and~$B$ can be established after negotiating a session key~$k$ through a Diffie-Hellman exchange using server~$S$ as a forwarder.
Each message exchanged between $A$ and~$B$ is encrypted with~$k$ by the sender, then the resulting ciphertext~$c$ is in turn encrypted with the sender's long-term authorization key (see \cref{sec:authorization}) and sent to~$S$.
Both layers use a similar encryption scheme, which we treat symbolically as a cryptographic primitive.
Upon receiving a message, $S$ uses the sender's authorization key, identified by 64-bit key fingerprint prepended to the message, to decipher the encrypted payload and recover the ciphertext~$c$, which is then encrypted again with the receiver's authorization key and forwarded to the receiver.

\ifreport
\begin{figure}[t]
  \begin{center}
    \includegraphics[width=0.6\columnwidth]{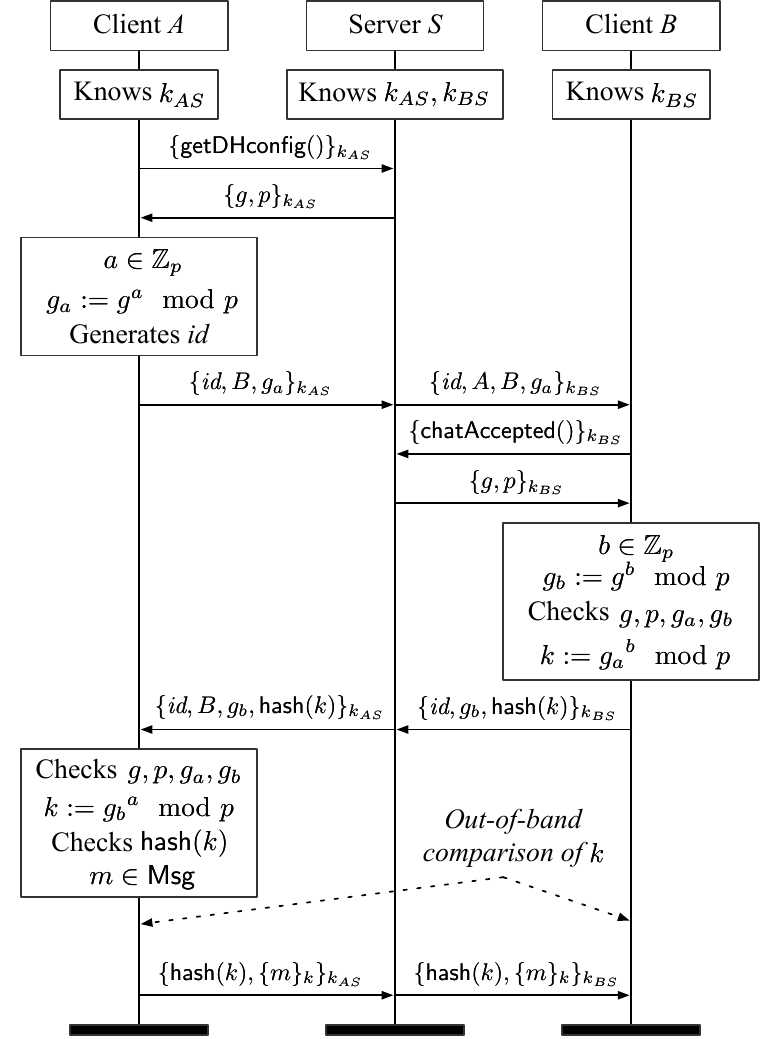}
    \caption{MTProto~2.0's protocol for secret chats.}
    \label{fig:secret-chat}
  \end{center}
\end{figure}
\fi

\ifreport
\subsection{Informal Description}
\label{sec:chat-informal}
The protocol, shown in \cref{fig:secret-chat},
\else
\bigskip
\noindent
\begin{minipage}[b]{0.45\textwidth}
\subsection{Informal Description}
\label{sec:chat-informal}
The protocol, shown aside,
\fi
is as follows:
\begin{enumerate}
  \item
  initiator~$A$ obtains DH parameters $(g,p)$ from~$S$, generates a random session identifier~\textit{id} and a half-key~$g_a$, and requests to start an encrypted chat with~$B$, including \textit{id} and~$g_a$ in the message;
  \item
  after $B$ has accepted the request, $B$ receives the DH parameters from~$S$ and computes a half-key~$g_b$, the session key~$k$ and a 64-bit fingerprint~$\HASH(k)$ of the key.
  The values $g_b$ and $\HASH(k)$ are then sent to~$A$, who can compute $k$ as well.
  The fingerprint~$\HASH(k)$ is not cryptographically strong: its stated purpose is only to prevent certain bugs in software implementations, especially during development.
\end{enumerate}
\ifreport
\else
\end{minipage}\hfill
    \includegraphics[width=.53\textwidth]{MTProto2-secret-chat}
\fi
This exchange is unauthenticated, so it is trivial for the server to act as a MitM and establish two different keys with~$A$ and~$B$, respectively.
To detect such attacks, after the DH exchange is completed, the clients are required to compare their respective key fingerprints\footnote{These are different from~$\HASH(k)$; in MTProto~2.0, they are 288-bit hashes that are typically displayed both as hexadecimal strings and QR-codes, suitable for visual comparison, by Telegram clients.} using a secure authenticated out-of-band channel.
Under this assumption, the protocol should guarantee the secrecy of the messages subsequently exchanged by $A$ and~$B$, encrypted with~$k$.
The clients are also supposed to perform suitable checks on the DH parameters, as described in \cref{sec:authorization}.
Note that, in actual implementations, the out-of-band check is left to users, i.e., it is not done automatically.

\subsection{Formalisation in ProVerif}
\label{sec:chat-formal}
Each (honest) client involved in the secret chat protocol is modelled as a distinct process (an initiator~$A$ and a responder~$B$), and communication happens via a common public channel (without
encrypting messages with authorization keys), which allows us to treat the server as an adversary equipped with the same knowledge as the server (essentially, all the authorization keys) and
implicitly performing message forwarding.
That allows the attacker to receive, manipulate, and resend the exchanged messages in the same way as the server could do, or impersonate a client if the clients do not perform the
required checks on the received parameters or on the generated key.

As the server controls the DH configuration used by the clients to derive their shared key, if the clients do not validate the obtained values then the server might be able to force both clients to use an easily guessable key.
For instance, if the server sends a subgroup generator equal to~$1$ and the clients do not perform any check, the derived session key will be~$1$ and the server will easily decrypt all the messages.

\looseness=-1
To model the out-of-band verification that users are asked to perform on a newly generated session key, we use a separate secure channel~$\tilde c$ available only to~$A$, $B$, and two auxiliary processes, one performing the required validation on behalf of the clients, and one skipping it:
\begin{align*}
&\LET\ \textsf{PerformOutOfBandKeyComparison}() =\\
&\quad\IN({\tilde c}, \textsf{QR}(X,Y,k));\\
&\quad\IN({\tilde c}, \textsf{QR}(=Y,=X,=k));\\
&\quad\EVENT\ \textsf{OutOfBandKeyComparisonSucceeded}(X, Y, k);\\
&\quad\EVENT\ \textsf{OutOfBandKeyComparisonSucceeded}(Y, X, k);\\
&\quad\OUT({\tilde c}, \textsf{QR}_{\textit{OK}}(X, Y, k));\\
&\quad\OUT({\tilde c}, \textsf{QR}_{\textit{OK}}(Y, X, k)).
\\[2ex]
&\LET\ \textsf{SkipOutOfBandKeyComparison}() =\\
&\quad\IN({\tilde c}, \textsf{QR}(X,Y,k));\\
&\quad\EVENT\ \textsf{OutOfBandKeyComparisonSkipped}(X,k);\\
&\quad\OUT({\tilde c}, \textsf{QR}_{\textit{OK}}(X,Y,k)).
\end{align*}
Each client~$X$ wishing to validate a session key negotiated with~$Y$ sends a private message $\textsf{QR}(X,Y,k)$ via~$\tilde c$ and waits for $\textsf{QR}_{\textit{OK}}(X,Y,k)$ as a confirmation.
The first of the two processes above waits for a $\textsf{QR}()$ message from a client and then expects a matching message from the other client.
Only when the second message arrives, the process proceeds by recording the event and sending out the confirmation messages.
The second process simply acknowledges a \textsf{QR}() message and records the corresponding event, without doing anything else.
Note that, in our model, out-of-band key validation implies mutual authentication.
How such channel is to be established in the real world, however, is not specified by the protocol.

Finally, we assume that honest clients behave as per the protocol,
except for the following deviations: a client may fail to verify its
DH parameters, as in the authorization protocol
(\cref{sec:auth-formal}), and a client may skip the out-of-band validation of the session key.

\subsection{Security properties verification}
\label{sec:chat-security}
\subparagraph{Secrecy.}
The main requirement of an end-to-end encrypted chat is, obviously,
secrecy: messages exchanged by~$A$ and~$B$ must be known only to~$A$
and~$B$. MTProto's secret chats guarantee secrecy conditional to the
strong assumption that clients do validate their keys through
a separate private channel. Formally, the secrecy query, which
ProVerif is able to prove, can be formulated as follows:
\begin{align*}
&\QUERY X\colon\PRINCIPAL, k\colon\SESSIONKEY, m\colon\MSG;\\
&\quad\ATTACKER(m)\\
&\quad\Rightarrow\EVENT(\textsf{OutOfBandKeyComparisonSkipped}(X,k))
\lor\EVENT(\textsf{ClientChecksDHConfig}(X,\bot)).
\end{align*}

\noindent In words, under the assumptions of our model, message~$m$ exchanged between two clients is kept secret unless one of the clients does not perform the mandatory checks of DH parameters or the clients omit the ``manual'' authentication of the shared key via an external secure channel.\footnote{Note that in our formalization (see \cref{sec:chat-formal}), it is not possible for one client to validate the session key and for the other to skip the validation.} Note that secrecy does not rely on the privacy of the authorization keys, i.e., the above query is true even if the authorization keys of the parties involved are leaked before the secret chat protocol starts. Secrecy, however, crucially relies on a step that requires active human interaction, at least in the current  implementations.

As for the confidentiality of the session key $k$, note that although in a correct run of the protocol $k$ remains secret (otherwise the attacker could obtain~$m$), $k$ is not indistinguishable from a random key: by comparing $\textsf{choice}[k,r]$ (i.e., one between the real key~$k$ and a random value~$r$) with the hash from the last message of the protocol, the server can tell apart $k$ from~$r$ with high probability. Accordingly, ProVerif cannot prove the corresponding observational equivalence.

\subparagraph{Integrity and authentication.}
The integrity and authenticity of a message exchanged by two clients during a secret chat session is preserved if the clients abide by the rules---in particular, if they perform the required out-of-band key validation in a way that implies mutual authentication.
But, even under such assumptions, the protocol remains vulnerable to a limited \emph{cross-session attack}.
The strongest result we could prove in ProVerif is the following:
\begin{align*}
&\QUERY i,i'\colon\textsf{ChatID},\; X,Y,I,R\colon\PRINCIPAL,\\
&\quad k\colon\SESSIONKEY,\; m\colon\textsf{Message};\\
&\INJEVENT(\textsf{ReceivesSecretChatMsg}(X,i,I,R,k,m))\\
&\Rightarrow\INJEVENT(\textsf{SendsSecretChatMsg}(Y,i',I,R,k,m))\\
&\lor\INJEVENT(\textsf{SendsSecretChatMsg}(Y,i',R,I,k,m))\\
&\lor\bigl(\EVENT(\textsf{OutOfBandKeyComparisonSkipped}(X,k))\\
&\quad\land\EVENT(\textsf{OutOfBandKeyComparisonSkipped}(Y,k))\bigr)\\
&\lor\bigl(\EVENT(\textsf{ClientChecksDHConfig}(X,\bot))\\
&\quad\land\EVENT(\textsf{ClientChecksDHConfig}(Y,\bot))\bigr).
\end{align*}
In this query, $I$ denotes the chat initiator and $R$ is the responder.
If the clients behave correctly, whenever an honest party $X$ (which, by the definition of the events, may be only $I$ or~$R$) receives a message~$m$ encrypted with the session key~$k$ in a chat session initiated by~$I$ with intended partner~$R$, exactly one message~$m$ encrypted with~$k$ has indeed been sent by~$Y$ (which, again, can only be $I$ or~$R$) in a session started by~$I$ with~$R$ \emph{or} in a parallel session started by~$R$ with~$I$.
The latter is due to the possibility that the server~$S$ impersonates the responder in two distinct sessions, one started by $X$ with~$Y$, and another started by $Y$ with~$X$, performing the following exchange:
\begin{description}
  \item
  [Session~1] $X\rightarrow S$: $g^x$
  \item
  [Session 2] $Y\rightarrow S$: $g^y$
  \item
  [Session 2] $S\rightarrow Y$: $g^x$
  \item
  [Session 1] $S\rightarrow X$: $g^y$
\end{description}
The server impersonating $Y$ in Session~1 and $X$ in Session~2 gets $X$ and~$Y$ to compute the same key $g^{xy}$ in both sessions. The server is then free to forward messages across such sessions: they are accepted because they are encrypted with a valid key. Note that this attack is limited to the same two parties $X$ and~$Y$ as long as the clients validate their keys.

In the query above we cannot require $i=i'$, because the server might forward $X$ and~$Y$ different chat \textit{id}'s. This does not seem to pose security risks, though: it is rather a correctness issue related to session management.
Anyway, a similar result additionally requiring $i=i'$ can be proved if the clients also compare their respective~\textit{id}'s during the out-of-band confirmation step, i.e., if $\textsf{QR}(i,X,Y,k)$ is sent instead of just~$\textsf{QR}(X,Y,k)$.

If the message is not sent by~$Y$, then $X$ and~$Y$ did not validate~$k$, or both clients skipped the needed checks on the DH parameters.
The former enables the server to perform the classical MitM attack on Diffie-Hellman.
The latter allows the server to mount a MitM attack similar to the one described in the last paragraph of \cref{sec:auth-security}.
Besides, it is necessary to assume that out-of-band validation
really implies mutual authentication: that is the case in our model,
but depending on how such validation is carried out in practice,
such an assumption may be too strong.
For instance, a malicious client $E$ could ``social-engineer'' two honest clients $A$ and~$B$ to make them believe that their key~$k$ match $E$'s (who does not have~$k$), e.g., $E$ might get the key fingerprint from $B$ and hand it to $A$ when asked by $A$ to compare their keys.

\section{Rekeying}
\label{sec:rekeying}
Keys used in secret chats are replaced every 100 messages or every week (if at least one message has been sent) using the protocol
\ifreport
shown in~\cref{fig:rekeying}.
\else
described below.
\fi
Old keys should be destroyed and never reused.
The exchange uses the same DH parameters obtained when the secret chat was first established (\cref{sec:secret-chat}).
Messages are transmitted through the secret channel already in place between the clients, so the server, who acts as a forwarder, can observe only the ciphertexts.
\subsection{Informal Description}
\label{sec:rekeying-informal}
The rekeying protocol is essentially a standard Diffie-Hellman exchange over a secure channel.
The claim that the channel is secure is only relative to the assumptions and results of \cref{sec:chat-security}.
Both clients possess the DH parameters~$(g,p)$ from their initial run of the secret chat protocol and a shared key~$k$.
The new shared key is derived as follows %
\ifreport
(see \cref{fig:rekeying}):
\begin{figure}[t]
  \begin{center}
    \includegraphics[width=0.6\columnwidth]{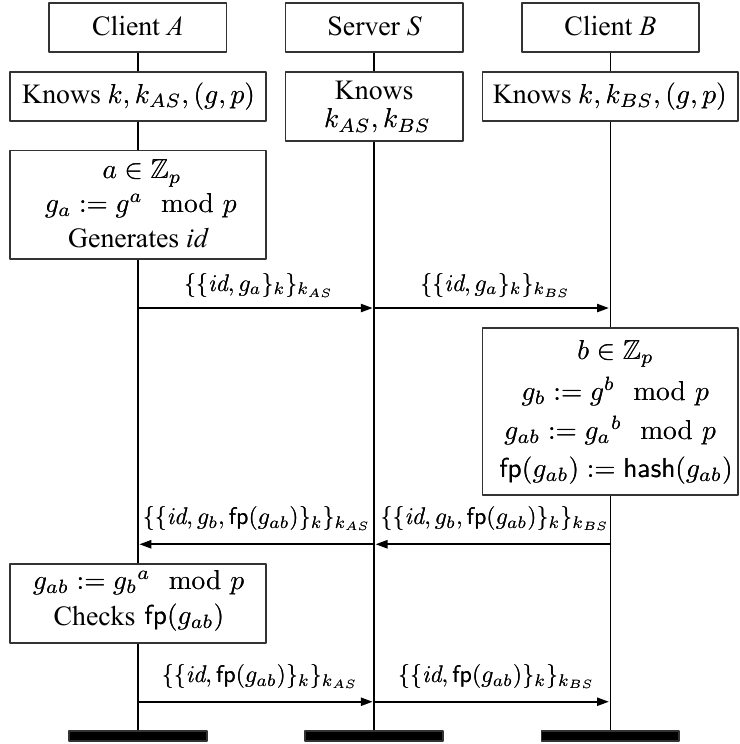}
  \end{center}
  \caption{The rekeying protocol.}
  \label{fig:rekeying}
\end{figure}
\else:
\bigskip
\noindent
\begin{minipage}[b]{0.45\textwidth}
\fi
\begin{enumerate}
  \item
  Client~$A$ generates a random session~\textit{id} and a random~$a\in\mathbb{Z}_p$, computes a half-key~$g_a$ and sends the pair~$(\textit{id},g_a)$ to~$B$.
  \item
  Client~$B$ generates a random $b\in\mathbb{Z}_p$ and computes its half-key~$g_b$ and the new shared key~$g_{ab}$.
  The half-key and a fingerprint of the shared key ($64$~bits of the SHA1 of the key) are sent to~$A$.
  \item
  Client~$A$ computes~$g_{ab}$, checks that the fingerprint of~$g_{ab}$ matches the received fingerprint, and sends the fingerprint back to~$B$ as an acknowledgment.
\end{enumerate}
The old key is no longer needed and can be deleted.
\ifreport
\else
\end{minipage}\hfill
\includegraphics[width=0.53\textwidth]{MTProto2-rekeying}
\fi

\subsection{Formalisation in ProVerif}
\label{sec:rekeying-formal}%
The formalisation of the rekeying protocol is similar to the formalization of the secret chat protocol, using a process for the initiator and one for the responder, while the server is untrusted and included in the attacker.
The computed fingerprints are not cryptographically strong and do not play any security role, but are meant only as a sanity check for the implementations.
The only important difference compared to a standard Diffie-Hellman exchange is that all the messages exchanged during rekeying are end-to-end encrypted with the current session key~$k$, which is assumed to be known only to the clients.

\subsection{Security properties verification}
\label{sec:rekeying-security}
\subparagraph{Secrecy and forward secrecy.}
Assuming that the current~$k$ is strong and secret, secrecy after rekeying is preserved even if the authorization keys of both parties are compromised before the rekeying protocol is executed.
This can be proved in ProVerif by running the rekeying protocol without encrypting the messages with the authorization
keys so that they are accessible to the adversary (which models the untrusted server), then letting $A$ and~$B$ exchange a message~$m$ encrypted with the new key, and finally verifying that the attacker cannot obtain~$m$: %
\begin{displaymath}
\QUERY\ \ATTACKER(m).
\end{displaymath}

Forward secrecy and future secrecy are guaranteed by the periodic rotation of the keys.
If an attacker recovers a session key, she can decrypt at most 100 messages or a week worth of messages.
While older or newer messages cannot be deciphered, in some circumstances such window of compromise might still be considered excessively wide.
Given the above, leaking an authorization key at any time does not compromise the secrecy of any message.

\vspace{-1ex}
\subparagraph{Authentication.}
If $A,B$ run the rekeying protocol negotiating a new key~$k$, then both $A$ and $B$, and only them, should know $k$.
Unfortunately, the rekeying protocol fails to provide \emph{implicit key confirmation} (the guarantee to~$A$ that $B$ can compute~$k$, and vice versa) and \emph{implicit key authentication} (the guarantee to~$A$ that no one but~$B$ knows~$k$, and vice versa)~\cite{Blake-Wilson:1999}.
A rogue client~$E$, who has already completed the secret chat protocol with $A$ and~$B$, may unfold the following attack:
\begin{itemize}
  \item
  $A$ starts rekeying with~$E$ and sends its half-key $g_a$;
  \item
  $E$ starts a parallel session with~$B$, sending $g_a$ to~$B$;
  \item
  $B$ sends $E$ its half-key~$g_b$;
  \item
  $E$ sends $A$ the half-key~$g_b$ obtained from~$B$.
\end{itemize}
At this point, $A$ and $B$ compute the same key, and each one mistakenly believes that the key is shared with~$E$.
The following query is disproved by ProVerif, even assuming that the DH parameters are good:
\begin{align*}
\QUERY&i\colon\textsf{SessionID},\; I,I',R,R'\colon\PRINCIPAL,k\colon\SESSIONKEY;\\
&\EVENT(\textsf{InitiatorNegotiatesNewKey}(i,I,R,k))\\
&\land\EVENT(\textsf{ResponderNegotiatesNewKey}(i,I',R',k))\\
&\Rightarrow I=I' \lor I=R' \lor R=I' \lor R=R'.
\end{align*}
If the initiator~$I$ accepts~$k$ from~$R$ and responder~$R'$ accepts~$k$ from~$I'$, ideally only two parties should be involved (the query is even weaker than that, asking only that at least one entity appears in both events).

As an example of how such a vulnerability could be exploited, consider the following \emph{game of spies}: Alice and Eve are agents from Eurasia, and Bob is an agent from Eastasia.
Alice trusts Eve and would never communicate with Bob, but Eve has
been paid by Bob to perform the above attack.
After the rekeying exchange described above, Bob would be able to transmit misleading information to Alice, who would accept it as coming from Eve.
The main advantage compared to having Eve act as a forwarder is that Bob does not need to rely on Eve in subsequent communication;
it is also a less risky strategy for Eve (she could eventually claim that her device has been hacked).

Telegram's documentation recommends that clients do not start two rekeying instances concurrently.
But the query above generates a trace involving four parties $A$, $B$, and two cooperating malicious clients $E,E'$, in which $A$ runs rekeying with~$E$, and $E'$ runs rekeying with~$B$.
Each client runs a single instance of the protocol.
So, the above recommendation is not a mitigation for this attack.

\section{Conclusions}\label{sec:concl}
We have presented the formalisation of the MTProto~2.0 protocol suite in the applied $\pi$-calculus, and its analysis using the protocol verifier ProVerif.
\ifreport
This approach adopts the symbolic Dolev-Yao threat model: an active intruder can intercept, modify, forward, drop, replay or reflect any message.
\fi
Within this model, we have provided a fully automated proof of the soundness of
MTProto~2.0's protocols for first authentication, normal chat, end-to-end encrypted chat,
and rekeying mechanisms with respect to several security properties,
including authentication, integrity, secrecy and perfect forward secrecy,
also in the presence of malicious servers and clients.
Moreover, we have discovered that the rekeying protocol is vulnerable to a theoretical \emph{unknown key-share} (UKS) attack \cite{Blake-Wilson:1999}: a malicious client $B$, with the help of another client~$E$, can induce a client $A$ to believe that she
(still) shares a secret key with $E$, and instead $A$ shares the key with~$B$.
The practical exploitability of this attack in actual implementations is still to be investigated.
Our formalization covers also the behaviour of the users, when relevant; e.g., if the users do not check the fingerprints of their shared keys, a MitM attack is possible.

Vulnerabilities may arise also from the cryptographic primitives, from implementation flaws (e.g.~insufficient checks), from side-channels exfiltration (such as timing or traffic analysis), or from incorrect user behaviour.
Hence, these aspects deserve further investigation and particular care in the implementation and use of this protocol.

In this work, the basic encryption primitive of MTProto~2.0 has been modeled as a perfect authenticated encryption scheme.
Although no attack on this scheme is known to date, in order to deem MTProto~2.0  secure we need to prove properties such as IND-CCA and INT-CTXT.
This proof cannot be done in a symbolic model like ProVerif's, but it can be achieved in a \emph{computational} model, using tools like CryptoVerif or EasyCrypt \cite{blanchet2007cryptoverif,barthe2013easycrypt}---which we leave to future work.
However, even if a flaw were found in the current encryption scheme,
the results in this paper would still be valid: the encryption
scheme could be replaced with a stronger one, and no other changes would be required.

Besides automatic tools like ProVerif and CryptoVerif, one may analyse (cryptographic) protocols in semi-automatic, interactive tools. A prominent example is EasyCrypt mentioned above, but there are also formalisations of the applied $\pi$-calculus or even general temporal logics in proof assistants like Coq or Isabelle/HOL (see, e.g.~\cite{DBLP:conf/cie/KahsaiM08,miculan:ic01,DBLP:conf/types/HonsellM95}).
In principle, these interactive tools allow us to formalize and prove any properties which can be proved ``on the paper'', but the burden on the user is greater than that induced by automatic tools.

Concerning implementation flaws, our formalisation can be used as a reference for the correct implementation of MTProto 2.0 clients (and servers).
Tools like Spi2Java or FS2PV can be useful to this end \cite{bhargavan2010modular,pozza2004spi2java}.
Also, particular attention must be paid to side-channel attacks, such as on timing or traffic analysis.
Another issue concerning the correct implementation of clients is that a server can craft malicious DH parameters, e.g., choosing generators that make discrete logarithms significantly easier to compute~\cite{Kobeissi:2017} or choosing non-primes that pass the 15-round Miller-Rabin test.
To prevent the first attack, MTProto prescribes that clients verify that the values received from the server are valid (see \cref{sec:auth-security}).
However, as far as we can see, MTProto 2.0 still suffers from the latter vulnerability.
A possible improvement is to require clients to check the proposed primes by means of deterministic primality algorithms \cite{agrawal2004primes,lenstra2019primality}.

\ifreport
\looseness=-1
Correct user behaviour is crucial in order to prevent MitM attacks in secret chats.
As we have seen, users must check the fingerprint of their authorization keys through an external safe channel; actually, this issue concerns not only MTProto~2.0 but
any protocol whose keys are defined by means of an insecure DH exchange. %
In practice, such checks are likely often ignored, or performed through the very same (supposedly secure) chat.
Hence, users seriously concerned about privacy must be educated about the correct procedure to follow.
\fi


\begin{thebibliography}{10}
	
	\bibitem{agrawal2004primes}
	Manindra Agrawal, Neeraj Kayal, and Nitin Saxena.
	\newblock Primes is in p.
	\newblock {\em Annals of mathematics}, pages 781--793, 2004.
	
	\bibitem{barthe2013easycrypt}
	Gilles Barthe, Fran{\c{c}}ois Dupressoir, Benjamin Gr{\'e}goire, C{\'e}sar
	Kunz, Benedikt Schmidt, and Pierre-Yves Strub.
	\newblock Easycrypt: A tutorial.
	\newblock In {\em Foundations of security analysis and design VII}, pages
	146--166. Springer, 2013.
	
	\bibitem{Bhargavan:2017}
	Karthikeyan Bhargavan, Bruno Blanchet, and Nadim Kobeissi.
	\newblock Verified models and reference implementations for the {TLS}~1.3
	standard candidate.
	\newblock In {\em IEEE European Symposium on Security and Privacy}, pages
	483--502, 2017.
	
	\bibitem{bhargavan2010modular}
	Karthikeyan Bhargavan, C{\'e}dric Fournet, and Andrew~D Gordon.
	\newblock Modular verification of security protocol code by typing.
	\newblock {\em ACM Sigplan Notices}, 45(1):445--456, 2010.
	
	\bibitem{Blake-Wilson:1999}
	Simon Blake-Wilson and Alfred Menezes.
	\newblock Unknown key-share attacks on the {S}tation-to-{S}tation ({STS})
	protocol.
	\newblock In {\em International Workshop on Public Key Cryptography}, pages
	154--170. Springer, 1999.
	
	\bibitem{blanchet2007cryptoverif}
	Bruno Blanchet.
	\newblock Cryptoverif: Computationally sound mechanized prover for
	cryptographic protocols.
	\newblock In {\em Dagstuhl seminar ``Formal Protocol Verification Applied''},
	volume 117, 2007.
	
	\bibitem{Blanchet:2016}
	Bruno Blanchet.
	\newblock Modeling and verifying security protocols with the {A}pplied {P}i
	{C}alculus and {P}ro{V}erif.
	\newblock {\em Foundations and Trends in Privacy and Security}, 1:1--135, 2016.
	
	\bibitem{cohn2017formal}
	Katriel Cohn-Gordon, Cas Cremers, Benjamin Dowling, Luke Garratt, and Douglas
	Stebila.
	\newblock A formal security analysis of the {S}ignal messaging protocol.
	\newblock In {\em IEEE European Symposium on Security and Privacy}, pages
	451--466, 2017.
	
	\bibitem{Dolev:1983}
	Danny Dolev and Andrew Yao.
	\newblock On the security of public key protocols.
	\newblock {\em IEEE Transactions on Information Theory}, 29(2):198--208, 1983.
	
	\bibitem{frosch2016secure}
	Tilman Frosch, Christian Mainka, Christoph Bader, Florian Bergsma, J{\"o}rg
	Schwenk, and Thorsten Holz.
	\newblock How secure is textsecure?
	\newblock In {\em 2016 IEEE European Symposium on Security and Privacy
		(EuroS\&P)}, pages 457--472. IEEE, 2016.
	
	\bibitem{DBLP:conf/types/HonsellM95}
	Furio Honsell and Marino Miculan.
	\newblock A natural deduction approach to dynamic logic.
	\newblock In Stefano Berardi and Mario Coppo, editors, {\em Types for Proofs
		and Programs, International Workshop TYPES'95, Selected Papers}, volume 1158
	of {\em Lecture Notes in Computer Science}, pages 165--182. Springer, 1995.
	
	\bibitem{Jakobsen:2015}
	Jakob Jakobsen.
	\newblock A practical cryptanalysis of the {T}elegram messaging protocol.
	\newblock Master's thesis, Aarhus University, September 2015.
	
	\bibitem{Jakobsen:2016}
	Jakob Jakobsen and Claudio Orlandi.
	\newblock On the {CCA} (in)security of {MTP}roto.
	\newblock In {\em Proceedings of the 6th Workshop on Security and Privacy in
		Smartphones and Mobile Devices}, pages 113--116, 2016.
	
	\bibitem{DBLP:conf/cie/KahsaiM08}
	Temesghen Kahsai and Marino Miculan.
	\newblock Implementing spi calculus using nominal techniques.
	\newblock In Arnold Beckmann, Costas Dimitracopoulos, and Benedikt L{\"{o}}we,
	editors, {\em Logic and Theory of Algorithms, 4th Conference on Computability
		in Europe, CiE 2008, Athens, Greece, June 15-20, 2008, Proceedings}, volume
	5028 of {\em Lecture Notes in Computer Science}, pages 294--305. Springer,
	2008.
	
	\bibitem{Kobeissi:2017}
	Nadim Kobeissi, Karthikeyan Bhargavan, and Bruno Blanchet.
	\newblock Automated verification for secure messaging protocols and their
	implementations: A symbolic and computational approach.
	\newblock In {\em 2017 IEEE European Symposium on Security and Privacy
		(EuroS\&P)}, pages 435--450. IEEE, 2017.
	
	\bibitem{lenstra2019primality}
	Hendrik~W Lenstra~Jr and Carl~B Pomerance.
	\newblock Primality testing with gaussian periods.
	\newblock {\em Journal of the European Mathematical Society}, 21(4):1229--1269,
	2019.
	
	\bibitem{miculan:ic01}
	Marino Miculan.
	\newblock On the formalization of the modal {\(\mathrm{\mu}\)}-calculus in the
	calculus of inductive constructions.
	\newblock {\em Inf. Comput.}, 164(1):199--231, 2001.
	
	\bibitem{pozza2004spi2java}
	Davide Pozza, Riccardo Sisto, and Luca Durante.
	\newblock Spi2java: Automatic cryptographic protocol {Java} code generation
	from spi-calculus.
	\newblock In {\em 18th International Conference on Advanced Information
		Networking and Applications. AINA 2004.}, volume~1, pages 400--405. IEEE,
	2004.
	
	\bibitem{Rad:2015}
	Alex Rad and Juliano Rizzo.
	\newblock A $2^{64}$ attack on {T}elegram, and why a super villain doesn't need
	it to read your telegram chats.
	\newblock
	\href{https://web.archive.org/web/20160425091011/http://www.alexrad.me/discourse/a-264-attack-on-telegram-and-why-a-super-villain-doesnt-need-it-to-read-your-telegram-chats.html}{http://www.alexrad.me},
	2015.
	\newblock Accessible via
	\href{https://web.archive.org}{https://web.archive.org}.
	
	\bibitem{rosler2018more}
	Paul R{\"o}sler, Christian Mainka, and J{\"o}rg Schwenk.
	\newblock More is less: on the end-to-end security of group chats in {S}ignal,
	{W}hats{A}pp, and {T}hreema.
	\newblock In {\em IEEE European Symposium on Security and Privacy}, pages
	415--429, 2018.
	
	\bibitem{Susanka:2017}
	Tom{\'a}{\v{s}} Su{\v{s}}{\'a}nka and Josef Koke{\v{s}}.
	\newblock Security analysis of the {T}elegram {IM}.
	\newblock In {\em Proceedings of the 1st Reversing and Offensive-Oriented
		Trends Symposium (ROOTS~2017)}, pages 1--8, 2017.
	
	\bibitem{Telegram-MTProto}
	Telegram.
	\newblock {MTP}roto mobile protocol.
	\newblock \url{https://core.telegram.org/mtproto/} (last accessed on February
	15, 2021), 2021.
	
	\bibitem{Telegram-Tech-FAQ}
	Telegram.
	\newblock Telegram {FAQ for the Technically Inclined}.
	\newblock \url{https://core.telegram.org/techfaq} (last accessed on April 15,
	2021), 2021.
	
\end{thebibliography}
\end{document}